# PNet: A Python Library for Petri Net Modeling and Simulation

Zhu En Chay

**Colossus Technologies LLP, Republic of Singapore**
*chayzhuen@colossus-tech.com*

Bing Feng Goh

**Singapore Institute of Technology, Republic of Singapore**
*gohbingfeng@gmail.com*

Maurice HT Ling

**Colossus Technologies LLP, Republic of Singapore**
**School of BioSciences, The University of Melbourne, Australia**
*mauriceling@colossus-tech.com*

**Abstract**

Petri Net is a formalism to describe changes between 2 or more states across discrete time and has been used to model many systems. We present PNet – a pure Python library for Petri Net modeling and simulation in Python programming language. The design of PNet focuses on reducing the learning curve needed to define a Petri Net by using a text-based language rather than programming constructs to define transition rules. Complex transition rules can be refined as regular Python functions. To demonstrate the simplicity of PNet, we present 2 examples – bread baking, and epidemiological models.

## 1. Introduction

Petri Nets are tools designed by C. A. Petri to model concurrent systems, as graphical representations and mathematical modeling tool for system of events [1-3]. The use of Petri Nets allows formal analysis of the model or process that is being depicted [1-3]. Petri Nets are populated by three types of objects – places, transitions and arcs [1-3]. A place is an input position that is connected to another transition via an arc. Place is usually depicted by circles and transitions as bars [1-3].

Petri Nets are often used for software engineering [4], system modeling [5] and even in biochemistry; such as biochemical reactions [6], signal transduction networks [7] and gene regulation networks [8]. An example of the use of Petri Nets is by Liu and Heiner [9] where they investigate biochemical reaction networks with the use of unifying Petri Net framework to model and analyze such networks [4]. This is because the properties of the processes can be studied: Terminating, Reachability, Safeness, Boundedness, Liveness, Reversibilty and Home State, Coverability, Persistence and Fairness [1]. Such properties can be studied [1] using Coverability tree method, Reachability Graphs and Incidence Matrix and State Equation.

There are libraries developed for modeling Petri Nets, such as SimForge GUI [10] incorporated within OpenModelica, and Petri Net Simulink Block (PNSB [11]) for MATLAB. SNAKES had been developed by Pommereau [12] as a library for implementing Petri Nets in Python programming language. SNAKES [12] adopts a high-level of object-oriented programming as tokens and all transition rules are implemented as Python objects. Although this provides flexibility, it may present as a steeper learning curve. It may be considerably more difficult to translate a text-based Petri Net specification into a model in SNAKES. At the same time, SNAKES [12] does not cater for complex transition rules that can only be implemented as a function. However, a strong advantage of SNAKES [12] is the incorporation of plugins, Petri Net analysis tools, and the ability to convert implemented Petri Nets into C language.

In this work, we present PNet an alternative pure Python library for Petri Net modeling by reducing its object-oriented programming overheads to its minimum, and adding Python functions as an alternative type of transition rule. Hence, PNet is likely to reduce the learning curve needed for a beginner to start experiencing Petri Nets before transitioning to more extensive library, such as SNAKES [12]. PNet has been incorporated into COPADS (https://github.com/mauriceling/copads), a library of algorithms and data structures, developed entirely in Python programming language and has no third-party dependencies. Future work aims to implement a GUI for improved usability and tools to analyze Petri Nets.

## 2. Description of PNet

In this section, we will describe PNet by using the steps required to write a simulation. There are 5 steps to writing a Petri Net simulation using PNet – establishing a Petri Net, adding places or states, adding transition rules, simulating the Petri Net, and generating the results file.

A Petri Net is established by importing PNet as a module and instantiating the PNet class within the imported module. This is followed by adding of places or states into the Petri Net using `add_places` method, which takes 2 parameters – the name of the place and a dictionary representing the initial tokens, where the dictionary key and value is the name of the token and the number of named tokens respectively. This allows for a place/state to have more than 1 type of tokens. For example, a mixed bowl of 100 red and green beans each can be stated as

```
net.add_places('bowl',
               {'red_beans': 100,
                'green_beans': 100})
```

However, there are situations whereby an infinite source or sink is needed; for example, the number of people to be born may be virtually infinite. In electronics, Earth is considered an infinite source of positive and negative charges. To cater for this need, a special place known as `ouroboros` (ouroboros is the name for the "infinity" symbol in mathematics) is defined with an infinite number of "U" tokens.

The third step is the addition of transition rule(s), using `add_rules` method. Each transition rules is named. Transitions are channels where the tokens move from a place to another, whereas the rules determine how the move occurs. Generally, a transition rule consists of a source place and destination place to direct the tokens, source token type for identification purpose and destination token type for getting a precise result. The processes of transitions rules will then be checked against the find value of tokens, using logical operators. The logical operators of the checks are determined by criterions, which are also known as the intended result after the going through the transitions.

There are 5 types of rules: step, ratio, delay, incubate, and function rules. The execution/firing of transition rules is time-step dependent. Although in the strictest sense, each rule should define only one transition; in practice, a single rule can trigger 1 or more transitions as it is possible to specify more than 1 transition in a rule. This can be seen as a syntactic shorthand provided by PNet.

Step rule based on a step-wise execution where the rule will be triggered at every time-step. The origin place and the token at the origin place have to be indicated; at the same time, the destination place and the affected token at the destination place must be specified. This represents a single transition. For example, given a bowl of red and green beans each, the following step rule defines the swapping of a bean at each time step,

```
net.add_rules('swap_bean, 'step',
  ['B1.red_bean -> B2.red_bean; 1',
   'B2.green_bean -> B1.green_bean; 1'])
```

Ratio rule is also a step-wise execution. Both step and ratio rules have similar parameters, with the difference of using the ratio of tokens to trigger the execution. The ratio that is intended for the transition is indicated, which will be check by a logical operator against the limit indicated. This can be used to define increasing or decreasing number of tokens moved. For example, given that there are 2 bowl (B1 and B2) where B1 contains all the red beans and B2 is an empty bowl, moving 10% of the remaining red beans in B1 to B2 can be defined as follow,

```
net.add_rules('swap_ratio', 'ratio',
    ['B1.red_beans -> B2.red_beans; 0.1; \
     B1.red_beans < 1; 0'])
```

Delay rule is a step rule with time interval between each token movement. In effect, delay rule can be used to produce a regular spiking movement. For example, moving 10 beans from bowl B1 to B2 once every $5^{th}$ time step, can be defined as

```
net.add_rules('interval_transfer', 'delay',
    ['B1.beans -> B2.beans; 10; 5'])
```

Incubation rule can be seen as a "do nothing for a period of time before a specific action". It requires a value and a timer, which has a logical check within to make sure that the conditions are met before sending to the destination place. For example, soaking a bowl of beans for 60 time steps (such as 60 minutes) once water is added, and transfer the soaked beans into a pot after soaking for 60 time steps, can be defined as

```
net.add_rules('soak', 'incubate',
    ['60; bowl.beans -> pot.beans; \
     bowl.water > 0'])
```

Function rule is a user-defined condition. Usually, function rules are used when the previous 4 rules do not fit the user's requirement. However, all forms of transition rules can be written as function rule; thus, function rule is the superset. Another common application of function rule is to change the type of tokens from one type to another. The variations between the transition rules are the conditions that trigger of the transition rules. The origin and destination place together with the initial token and the final token are required for the computation. For example, the above ratio rule

```
net.add_rules('swap_ratio', 'ratio',
    ['B1.red_beans -> B2.red_beans; 0.1; \
     B1.red_beans < 1; 0'])
```

can be written as the following function rule,

```
def bean_swap(places):
    place = places['B1']
    n = place.attributes['red_beans']
    if n > 0.0:
        return 0.0
    else:
        return 0.1 * n
net.add_rules('swap_ratio', 'function',
    ['B1.red_beans -> B2.red_beans',
     bean_swap, 'B1.red_beans > 0'])
```

Function(s) to be used in function rule(s) takes only one parameter, `places`, which is the dictionary of states/places in PNet. Each state/place can be accessed using the name of the state/place as key to the `places` dictionary. Tokens linked to a particular state/place are implemented as an `attributes` dictionary and be accessed using the name of token.

After rules definition, the fourth step is to simulate the Petri Net using `simulate` or `simulate_yield` method. The rules mentioned above will then be executed, by setting the wall time of current simulation and the interval. The wall time will be check against the rules to ensure that it fulfills the time interval indicated.

The `simulate` method will stores the simulation results in the memory space; thus, increased intervals of reporting causes more reports to be generated, which causes memory to run out at a faster rate. On the other hand, `simulate_yield` method is a generator function, which does not pre-store all the simulation results in memory. The parameters of `simulate` method are length of time to simulate, time step, and reporting frequency. However, `simulate_yield` method only requires length of time to simulate, and time step.

Finally, PNet provides a method to process the simulation results into a format suitable for CSV file output. The reports will be generated with each step count of current simulation, in the memory of each token status. The status of the tokens can be also report by generating a list representing the status from one step or the entire simulation.

Simulation and reporting are usually related to each other. For example, the following code snippet demonstrates the simulation and report generation using both `simulate` or `simulate_yield` method:

```
length_of_simulation = 100
timestep = 1
report_frequency = 1
# for simulate method
net.simulate(length_of_simulation,
             timestep, report_frequency)
status = net.report_tokens()

# for simulate_yield method
status = [d for d in
  net.simulate_yield(length_of_simulation,
                     timestep)]
status = [(d[0], net.report_tokens(d[1]))
          for d in status]
```

## 3. Examples

Two examples are presented to illustrate the use of PNet. The first example is a light-hearted example of bread baking while the second example is a more serious but simple model of epidemiology.

**Example 1: Bread Baking.** In this example, a bread baking recipe was modeled (see Appendix A for implementation) and simulated for 90 time steps. It is worth noting that this recipe utilized all features of PNet except the use of infinite tokens from Ouroboros. This recipe calls for 1000 g of flour, 500 g of water, 20 g of sugar, and 1 g of yeast in the following steps:

1. Turn on the mixer and add 100 g of flour, 50 g of water, and 2 g of sugar at each time step. Add 0.5 g of yeast into mixer with 5 time step interval in between each addition.
2. In each time step, the mixer will turn 80 g of flour, 40 g of water, 1.5 g of sugar, and 1 g of yeast into dough.
3. After mixing is completed, leave dough to rise in mixer for 10 time steps.
4. Transfer dough into pan, and leave dough to rise in pan for another 10 time steps.
5. Bake at 400ºC. In each time step, 30% of the remaining dough will be baked into bread. Baking is completed when there is less than 1 g of dough remaining.
6. Transfer the bread to the table for cooling. In each time step, 10% of the heat will dissipate until room temperature of 30ºC is reached.
7. Enjoy your bread.

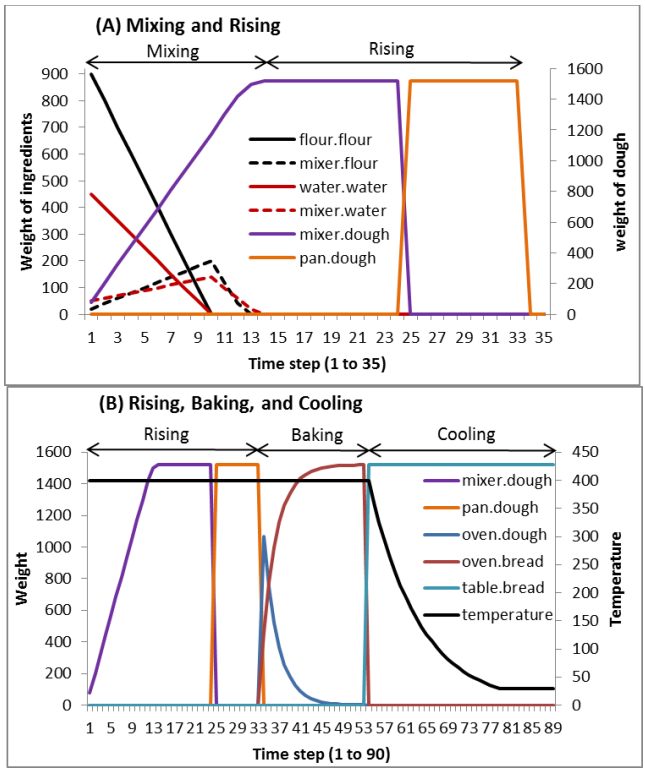

Fig. 1 Token Values in Bread Baking Simulation. Graph A shows the mixing of ingredients into bread dough and the rising of the dough. Graph B re-illustrate the rising of the dough and continue through the baking and cooling process.

Steps 1 and 2 are the addition and mixing the ingredients into bread dough. The rate of dough formation is slower than the rate of ingredients addition; for example, 100 g of flour and 50 g of water are added into the mixer per time step but only 80 g of flour and 40 g of water are converted to dough. This results in the accumulation of 20 g of flour and 10 g of water in the mixer until all flour and water are added (as seen in the mixing stage in Figure 1A); after which, the accumulated flour in the mixer is converted into bread dough. Once all ingredients are mixed into bread dough, it undergoes 2 stages of rising (Steps 3 and 4). After which, the dough is being baked at 400°C (Step 5). The baking process is represented as transferal of 30% of the dough tokens into bread tokens at each time step. Once the bread is baked as represented by negligible remains of dough (less than 1 g), the bread is transferred to a table and cooled (Step 6). The cooling from 400°C to room temperature of 30°C is represented by another function.

**Example 2: Epidemiological Models.** Epidemiological Models are frameworks of ecological and epidemiological phenomena that are often used to study interactions between the host and the pathogen [13]. Epidemiological models have proven useful for the study of evolutionary dynamics of evolutionary dynamics and predicting properties of the spread of pathogen like prevalence and duration [13]. Alphabet models are frameworks of a population whereby susceptible individuals are considered to be invaded by an infectious agent [13]. The population is divided into three epidemiological subclasses: S denotes susceptible to diseases, I denotes number of infected individuals and R denotes number of individuals who at time no longer contribute to spread of diseases [13].

The Susceptible-Infectious-Susceptible (SIS) model is predicated on the pathogen infects susceptible humans, resulting in an infection and recovers from infection and returns back to the susceptible class again [14]. Infectious hosts recover at a constant per capita rate, $\gamma$ and $\beta$ is the rate of infection of the susceptible class [14]. SIS model is for fast evolving virus and infections that do not provide immunity [14]. The Susceptible-Infectious-Recovered (SIR) model is similar to SIS model except that the pathogen leads to lifelong immunity [15]. The individuals who were infected and recovered from infection are immune to reinfection, possessing lifelong immunity [15]. This model is used for viral diseases such as measles, mumps and rubella [14]. The Susceptible-Infectious-Recovered-Susceptible (SIRS) model is similar to that of SIR model, except that the immunity that are acquired is temporary [16]. The individuals who were infected and recovered from infection are not immune to reinfection [14]. Example of such infection modeled by SIRS is tuberculosis [16].

However, it is common for most epidemiological models to be implemented as a system of ordinary differential equations (ODEs) [17-20]. ODEs and Petri Net are two of the most common mathematical constructs for mathematical modeling [21]. Hence, a method to represent an ODE in the form of Petri Net notation is needed and the correspondence between ODE and state-transition network is provided by Soliman and Heiner [22]. Briefly, an ODE models the change of a state over time while Petri Net models the movement which results in the change of state over time (Figure 2). In the context of states (nodes) and transitions (arcs), this suggests that ODEs models the nodes while Petri Net models the transitions, which gives rise to an easy conversion between ODE representations and Petri Net representations, assuming that the unit for time is the same under both representations.

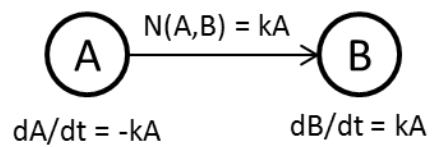

Fig. 2 Correspondence between Ordinary Differential Equation and Petri Net Transition Rule.

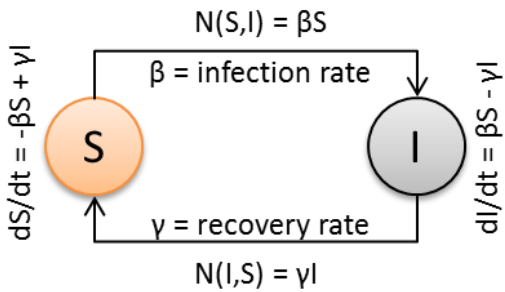

Fig. 3 SIS Epidemiological Model.

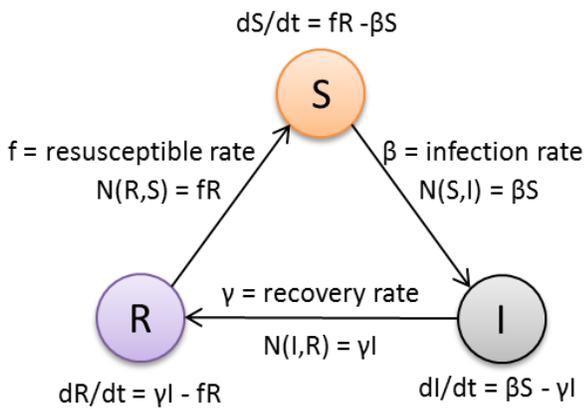

Fig. 4 SIR/SIRS Epidemiological Model.

Hence, the standard system of ODEs for SIRS [17, 18] can be readily converted into Petri Net representations (Figures 3 and 4). The implementation of SIRS model using PNet is given as Appendix B.

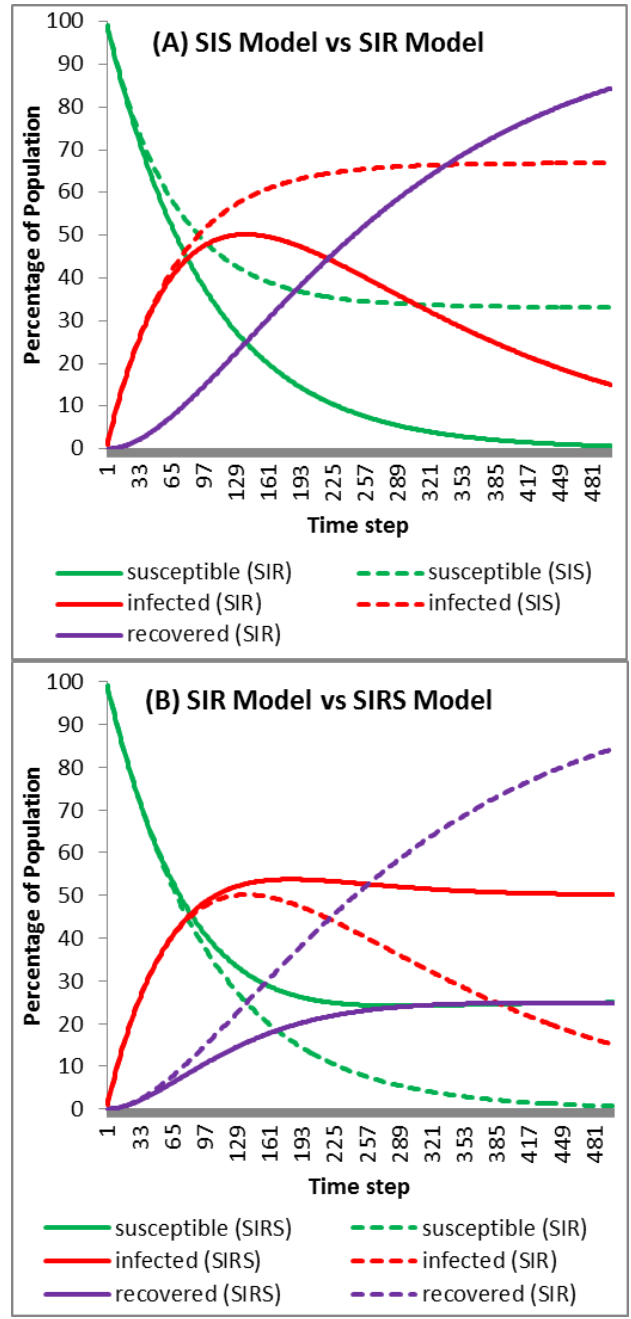

Fig. 5 Schematics of a Data Frame.

Our simulation results show that the proportion of infected population and susceptible population reaches equilibrium over time (Figure 5a). As there is also no immunity conferred after recovery, it is expected to have a constant infected population [23], also known as endemic population. This is under the assumption that there are no birth and death for the entire duration, and the disease is not death causing. When there is immunity as a result of recovery, SIS model becomes SIR model where the population gradually becomes fully immune if there is no additional birth (Figure 5b). This is similar to the case of chickenpox [24], which confers lifelong immunity to most recovered patients, leading to children being most susceptible to chicken pox and most adults immune. However, if the conferred immunity is temporary, re-infection is possible and this leads to SIRS model from SIR model (Figure 5b). Our results show that SIRS model behaves in similar manner compared to SIS model where there is a constant pool of infected (endemic) individuals [25]. In spite of this, there is also a constant pool of immune individuals whom had recently recovered from the disease. This is expected when the infection agent can re-infect a recovered person.

## Acknowledgement

The authors wish to thank HJ Wang (Nanyang Technological University) for valuable discussions during the developmental and testing phases.

## References

[1] W. Reisig, "Petri Nets: An Introduction", Springer–Verlag, 1985.
[2] W. Reisig, "Understanding Petri Nets: Modeling Techniques, Analysis Methods, Case Studies", Springer–Verlag, 2013.
[3] T. Murata, "Petri Nets: Properties, Analysis and Applications", Proceedings of the IEEE 77, 1989, 541-580.
[4] A. Bobbio, "System Modelling with Petri Nets". Instituto Elettrotecnico Nazionale Galileo Ferraris Strada Delle Cacce 91, 10135 Torino, Italy, 1990, 14-15.
[5] S. Hardy and R. Iyengar, "Analysis of Dynamical Models of Signaling Networks with Petri Nets and Dynamic Graphs", Modeling in Systems Biology 16, 2011, 225-251.
[6] D. Gilbert and M. Heiner M, "From Petri nets to differential equations–an integrative approach for biochemical network analysis", Petri Nets and Other Models of Concurrency-ICATPN 2006, 2006, 181-200.
[7] C. Chaouiya, "Petri Net Modelling of Biological Networks", Briefings in Bioinformatics 8, 2007, pp 210-219.
[8] H. Matsuno and A. Doi, "Hybrid Petri Net Representation of Gene Regulatory Network", Pacific Symposium on Biocomputing 5, 2000, 338-349.
[9] F. Liu and M. Heiner, "Modeling Membrane Systems using Colored Stochastic Petri Nets", Natural Computing 12, 2013, 617-629.
[10] S. Proß, B. Bachmann, R. Hofestädt, K. Niehaus, R. Ueckerdt, F.-J. Vorhölter, P. Lutter, A. Stadtholz, "Modeling a Bacterium's Life: A Petri-Net Library in Modelica", Conference Proceedings of Modelica 2009, 2009.
[11] M. Matcovschi, C. Popescu, and O. Pastravanu, A new approach to hybrid system simulation: Development of a simulink library for petri net models," Journal of Control Engineering and Applied Informatics 7, 2005, 55-62.
[12] F. Pommereau, "SNAKES: A Flexible High-Level Petri Nets Library (Tool Paper)", Proceedings of the 36th International Conference on Petri Nets (PETRI NETS 2015), 2015, 254-265.
[13] F. Brauer, "Compartmental Models in Epidemiology", In Lecture Notes in Mathematics 1945, 2008, 19-79.
[14] F. Brauer and C. Castillo-Chavez, "Basic Models in Epidemiology", In Ecological Time Series 1995, 410 – 447.
[15] M. J. Keeling and K. T.D. Eames, "Networks and epidemic models", Journal of the Royal Society Interface 2, 2005, 295-307.
[16] C. Ozcaglar, A. Shabbeer, S. L. Vandenberg, B. Yener, K. P. Bennett, "Epidemiological models of *Mycobacterium Tuberculosis* complex infections", Mathematical Biosciences 236, 2012, 77 – 96.
[17] P. Munz, I. Hudea, J. Imad and R. J. Smith, "When Zombies Attack!: Mathematical Modelling of Outbreak of Zombie Infection", Infectious Disease Modelling Research Progress, 4, 2009, 133-150.
[18] M. Ling, "COPADS IV: Fixed Time-Step ODE Solvers for a System of Equations Implemented as a Set of Python Functions", Advances in Computer Sciences 5, 2016, xx-xxx.
[19] J.P. Aparicio and M. Pascual, "Building Epidemiological Models from R(0): An Implicit Treatment of Transmission in Networks", Proceedings of the Royal Society B: Biological Sciences 274, 2007, 505-512.
[20] L. Kong, J. Wang, W. Han and Z. Cao, "Modeling Heterogeneity in Direct Infectious Disease Transmission in a Compartmental Model", International Journal of Environmental Research and Public Health 13, 2016, 253.
[21] M. Ling, "Of (Biological) Models and Simulations", MOJ Proteomics & Bioinformatics 3, 2016, 00093.
[22] S. Soliman and M. Heiner M, "A Unique Transformation from Ordinary Differential Equations to Reaction Networks", PLoS One 5, 2010, Article e14284.
[23] H.W. Hethcote and P. van den Driessche, "An SIS Epidemic Model with Variable Population Size and a Delay", Journal of Mathematical Biology 34, 1995, 177-194.
[24] "Facts about chickenpox", Paediatrics & Child Health 10, 2005, 413-414.
[25] Y. Chen, J. Yang and F. Zhang, "The Global Stability of an SIRS Model with Infection Age", Mathematical Biosciences and Engineering 11, 2014, 449-469.

## Appendix A: Code for Bread Baking

```python
import pnet

net = pnet.PNet()

# The ingredients
net.add_places('flour', {'flour': 1000})
net.add_places('water', {'water': 500})
net.add_places('sugar', {'sugar': 20})
net.add_places('yeast', {'yeast': 1})

# The "utensils"
net.add_places('mixer', {'flour': 0,
    'water': 0, 'sugar': 0,
    'yeast': 0, 'dough': 0})
net.add_places('pan', {'dough': 0})
net.add_places('oven', {'dough': 0,
    'bread': 0})
```

```python
net.add_places('table', {'bread': 0,
    'temperature': 400})
net.add_places('air', {'heat': 0})

# The steps
net.add_rules('add_flour', 'step',
    ['flour.flour -> mixer.flour; 100'])
net.add_rules('add_water', 'step',
    ['water.water -> mixer.water; 50'])
net.add_rules('add_sugar', 'step',
    ['sugar.sugar -> mixer.sugar; 2'])
net.add_rules('add_yeast', 'delay',
    ['yeast.yeast -> mixer.yeast; 0.5; 5'])
net.add_rules('blend', 'step',
    ['mixer.flour -> mixer.dough; 80',
     'mixer.water -> mixer.dough; 40',
     'mixer.sugar -> mixer.dough; 1.5',
     'mixer.yeast -> mixer.dough; 1'])
net.add_rules('rise', 'incubate',
    ['10; mixer.dough -> pan.dough; \
     mixer.flour == 0; mixer.water == 0; \
     mixer.sugar == 0; mixer.yeast == 0'])
net.add_rules('set', 'incubate',
    ['10; pan.dough -> oven.dough; \
     pan.dough > 0'])
net.add_rules('bake', 'ratio',
    ['oven.dough -> oven.bread; 0.3; \
     oven.dough < 1; 0'])
net.add_rules('transfer', 'incubate',
    ['1; oven.bread -> table.bread; \
     oven.dough == 0'])
def cooling(places):
    place = places['table']
    temp = place.attributes['temperature']
    if temp <= 30.0: return 0.0
    else: return 0.1 * temp
net.add_rules('cool', 'function',
    ['table.temperature -> air.heat',
     cooling, 'table.bread > 0'])

# Bake the bread !!!
net.simulate(90, 1, 1)

# Generate results file
data = net.report_tokens()
headers = ['timestep'] + data[0][1]

f = open('bread.csv', 'w')
f.write(','.join(headers) + '\n')
for tdata in data:
    tdata = [tdata[0]] + \
            [str(x) for x in tdata[2]]
    f.write(','.join(tdata) + '\n')
f.close()
```

## Appendix B: Code for SIRS Model

```python
import pnet

infection = 0.01
recovery = 0.005
resusceptible = 0.01

net = pnet.PNet()

net.add_places('susceptible',
    {'susceptible': 100})
net.add_places('infected',
    {'infected': 0})
net.add_places('recovered',
    {'recovered': 0})

def susceptible_infected(places):
    place = places['susceptible']
    susceptible = \
        place.attributes['susceptible']
    return infection * susceptible

def infected_recovered(places):
    place = places['infected']
    infected = place.attributes['infected']
    return recovery * infected

def recovered_susceptible(places):
    place = places['recovered']
    recovered = \
        place.attributes['recovered']
    return resusceptible * recovered

net.add_rules('infection', 'function',
    ['susceptible.susceptible -> \
     infected.infected',
     susceptible_infected,
     'susceptible.susceptible > 0'])
net.add_rules('recovery', 'function',
    ['infected.infected -> \
     recovered.recovered',
     infected_recovered,
     'infected.infected > 0'])
net.add_rules('resusceptible', 'function',
    ['recovered.recovered -> \
     susceptible.susceptible',
     recovered_susceptible,
     'recovered.recovered > 0'])

net.simulate(500, 1, 1)

data = net.report_tokens()
headers = ['timestep'] + data[0][1]

f = open('sirs.csv', 'w')
f.write(','.join(headers) + '\n')
for tdata in data:
    tdata = [tdata[0]] + [str(x) for x in tdata[2]]
    f.write(','.join(tdata) + '\n')
f.close()
```